\documentclass[pre,twocolumn,aps]{revtex4-1}

\usepackage{amsmath}
\usepackage{graphicx}
\usepackage{color}

\newcommand{\be}{\begin{equation}}
\newcommand{\ee}{\end{equation}}

\newcommand{\benn}{\begin{equation*}}
\newcommand{\eenn}{\end{equation*}}

\newcommand{\bea}{\begin{eqnarray}}
\newcommand{\eea}{\end{eqnarray}}

\newcommand{\mZ}{\mathcal{Z}}

\newcommand{\eb}[1]{{\color{black} #1}}
\newcommand{\colorg}[1]{{\color{black} #1}}

\usepackage{bm}

\begin{document}

\title{Participation ratio for constraint-driven condensation with superextensive mass}

\author{Giacomo Gradenigo}
\affiliation{LIPHY, Universit\'e Grenoble Alpes and CNRS, F-38000 Grenoble, France; ggradenigo@gmail.com}

\author{Eric Bertin}
\affiliation{LIPHY, Universit\'e Grenoble Alpes and CNRS, F-38000 Grenoble, France; eric.bertin@univ-grenoble-alpes.fr}

\begin{abstract}

  Broadly distributed random variables with a power-law distribution
  $f(m) \sim m^{-(1+\alpha)}$ are known to generate condensation
  effects, in the sense that, when the exponent $\alpha$ lies in a
  certain interval, the largest variable in a sum of $N$ (independent
  and identically distributed) terms is for large $N$ of the same
  order as the sum itself. \colorg{In particular, when the
    distribution has infinite mean ($0<\alpha<1$) one finds
    \emph{unconstrained} condensation, whereas for $\alpha>1$
    \emph{constrained} condensation takes places fixing the total mass
    to a large enough value $M=\sum_{i=1}^N m_i > M_c$.} \colorg{In
    both cases,} a standard indicator of the condensation phenomenon
  is the participation ratio $Y_k=\langle \sum_i m_i^k / (\sum_i
  m_i)^k\rangle$ ($k>1$), which takes a finite value for $N \to
  \infty$ when condensation occurs.  To better understand the
  connection between \colorg{constrained and unconstrained
    condensation}, we study here \colorg{the situation} when the total
  mass is fixed to a superextensive value $M \sim N^{1+\delta}$
  ($\delta >0$), hence interpolating between the unconstrained
  \colorg{condensation} case (where the typical value of the total
  mass scales as $M\sim N^{1/\alpha}$ for $\alpha<1$) and the
  extensive constrained mass. \colorg{In particular we show that for
    exponents $\alpha<1$ a condensate phase for values $\delta >
    \delta_c=1/\alpha-1$ is separated from a homogeneous phase at
    $\delta < \delta_c$ from a transition line, $\delta=\delta_c$,
    where a \emph{weak} condensation phenomenon takes place.} We focus
  on the evaluation of the participation ratio as a generic indicator
  of condensation, also recalling or presenting results in the
  standard cases of unconstrained mass and of fixed extensive mass.
\end{abstract}

\maketitle


\section{Introduction}
\label{intro}


In the context of the sum of a large number of positive random
variables, \eb{an interesting phenomenon occurs when a single variable
  carries a finite fraction of the sum \cite{SEM17}.  Such a
  phenomenon has been put forward for instance in the context of the
  glass transition \cite{MPV87,Bouchaud97}.  In the framework of
  particle or mass transport models
  \eb{\cite{BBJ97,MKB98,GSS03,MEZ05,EMZ06,EH05,S08-leshouches,HMS09,WCBE14,EW14}},
  where the sum of the random variables is fixed to a constant value
  due to a conservation law of the underlying dynamics, this
  phenomenon has been called ``condensation''. This condensation
  phenomenon has since then been reported in different contexts like
  in extreme value statistics \cite{EM08}, and in the sample variance
  of exponentially distributed random variables as well as} for
conditioned random-walks \cite{SEM14,SEM14b,SEM17}. A similar
mechanism is also at the basis of the condensation observed in the
non-equilibrium dynamics of non-interacting field-theoretical models
\cite{ZCG14,CGP15,Z15,CSZ17}.  A more general type of condensation,
induced by interaction, has also been put forward \cite{EHM06}, but in
the following we shall focus on cases without interaction, apart from
a possible constraint on the total mass.

\eb{As mentioned above, standard condensation results from the presence of a constraint fixing the sum of the random variables to a given value.
However, the fact that a single random variable carries a finite fraction of the sum is also observed for fat-tailed random variables with infinite mean ---a phenomenon sometimes called the Noah effect \cite{MW68,Magd08}.
The goal of the present paper is to present a comparative study of
these two scenarios, that we shall respectively denote as {\em constrained condensation} and {\em unconstrained condensation}. Note that the term 'condensation'
is usually used in the literature to describe the constrained case, but we shall extend its use to the unconstrained case, to emphasize possible analogies between the two scenarios.}
Considering the set of
$N$ random variables $m_i$ with joint distribution
$P(m_1,\ldots,m_N)$, \eb{unconstrained} condensation takes place
when the sum $M=\sum_{i=1}^N m_i$, in the limit $N\rightarrow \infty$,
is dominated by few terms, i.e., a number of terms of order
$\mathcal{O}(1)$. This happens for instance to the sum of $N$
independent and identically distributed (iid) Levy-type random
variables, with probability density $f(m)$ such that $f(m) \approx
A/m^{1+\alpha}$ when $m \to \infty$, with an exponent
$0<\alpha<1$. This \eb{unconstrained} condensation effect (sometimes also referred to as 'localization' \cite{Bouchaud03} \eb{depending on the context}) is often characterized by the
participation ratio $Y_k$ \cite{Bouchaud97,Derrida97}, defined as
\be 
Y_k = \left< \frac{\sum_{i=1}^N m_i^k}{\left( \sum_{i=1}^N m_i\right)^k} \right>,
\label{eq:part-ratios}
\ee
where $k>1$ is a real number, and where the brackets indicate an average over the $m_i$'s. For broadly distributed random variables
it can be shown, with the calculation presented in~\cite{Derrida97}
and briefly recalled here in Sec.~\ref{unconstrained}, that there is a
critical value $\alpha_c=1$ for the exponent of the
power-law distribution such that for $\alpha > 1$ the asymptotic value
of the participation ratio is zero, $\lim_{N\rightarrow\infty} Y_k =0
$, whereas for a broad enough tail, $0< \alpha < 1$, one has
$\lim_{N\rightarrow\infty} Y_k > 0$ for any value $k>1$. The average
in Eq.~(\ref{eq:part-ratios}) is computed with respect to the
probability distribution
$P(m_1,\ldots,m_N)=\prod_{i=1}^N f(m_i)$.  The participation ratio is
therefore the `order parameter' for condensation in the sum of random
variables. It is in fact easy to see from Eq.~(\ref{eq:part-ratios})
that when all the random variables contribute `democratically' to the
sum, namely when each of them is of order $m_i\sim 1/N $, then the
asymptotic behaviour of the participation ratio is $Y_k \sim
1/N^{k-1}$, which goes to zero when $N \to \infty$. In contrast, if
the sum is dominated by few terms of order $m_i \sim N$,
asymptotically one has $Y_k \sim 1$.

As a physical example, the relevance of participation ratios to unveil
\eb{unconstrained} condensation in the sum of broadly distributed random variables was
also shown for the condensation in phase space associated to the glass
transition in the Random Energy Model
(REM)~\cite{Derrida97,Bouchaud97}. The REM is a system with $2^N$
configurations, where each configuration $i$ has the Boltzmann weight
$e^{-\beta E_i}$, and the energies $E_i$ are iid random variables,
usually assumed to have a Gaussian distribution with a variance
proportional to $N$. The random variables with respect to which
the glass phase corresponds to a condensed phase are the
probabilities $z_i(\beta) = e^{-\beta E_i}$ of the different
configurations.
The corresponding participation ratio takes the same form as
Eq.~(\ref{eq:part-ratios}), simply replacing $m_i$ by $z_i(\beta)$.
It has been shown \cite{Derrida97,Bouchaud97} that for values of the
inverse temperature $\beta>\beta_c$, where $\beta_c$ is the critical
value of the glass transition, the value of the sum
$\mathcal{Z}=\sum_{i=1}^{2^N} z_i(\beta)$ is dominated in the limit
$N \rightarrow\infty$ by $\mathcal{O}(1)$ terms: in this case the asymptotic value
of $Y_k$ is finite. In particular one can prove that for an inverse
temperature $\beta > \beta_c$ the participation ratio of the REM has
precisely the same form as for the sum of iid Levy random variables
$m_i$ with distribution $p(m_i) = m_i^{1+\beta_c/\beta}$ (the exponent
$\alpha=\beta_c/\beta$ is thus proportional to temperature).

In the above cases, with iid random variables, \eb{unconstrained} condensation occurs for
$\alpha <1$, that is when the first moment of the power-law
distribution is infinite.  The situation is different, though, when
one considers power-law distributed random variables with a fixed
total sum $M=\sum_{i=1}^N m_i$, a case which we refer to as
\emph{constraint-driven} condensation, \eb{or simply {\em constrained condensation}}. Such a phenomenon,
\eb{which is also related to the large of heavy-tailed sums
(see, e.g., \cite{MN98})},
is found for instance in the stationary distribution of the discrete
  Zero Range Process and its continuous variables
  generalization~\cite{MEZ05,EMZ06,S08-leshouches}. The latter is
  represented by a lattice with $N$ sites, each carrying a continuous
  mass $m_i$, endowed with some total-mass conserving dynamical
  rules. For this model the stationary distribution is:
\be 
P(m_1,\ldots,m_N|\rho) = \frac{1}{\mZ_N(\rho)} \prod_{i=1}^N f(m_i)~\delta\left[ \rho N -\sum_{i=1}^N m_i \right],
\label{eq:constrained-prob}
\ee
where $\rho$ is the average density fixed by the initial total mass $M = \rho N$, and where
\be 
\mZ_N(\rho) = \int_0^{\infty} dm_1 \ldots dm_N ~\prod_{i=1}^N
f(m_i)~\delta\left[ \rho N -\sum_{i=1}^N m_i \right].  
\label{eq:canonical-partition}
\ee
is a normalization constant (or partition function).
In mass transport models the shape of the distribution
$f(m_i)$ depends on the dynamical rules, and has typically a power-law tail,
\be 
f_\alpha(m) \approx \frac{A}{m^{1+\alpha}} .
\label{eq:local-dist}
\ee
In~\cite{MEZ05,EMZ06,S08-leshouches} it has been shown that, in the
presence of a constraint on the total value of the mass,
\eb{constrained} condensation never takes place for exponents of the
local power-law distribution in the interval $0<\alpha<1$, while on
the contrary when $\alpha > 1$ there exists a critical value $\rho_c$
such that for $\rho > \rho_c$ the system is in the condensed phase.
It thus turns out that constraining the random variables to have a
fixed sum deeply modifies their statistical properties in this case
---while naive intuition based on elementary statistical physics like
the equivalence of ensembles may suggest that fixing the sum may not
make an important difference.  \eb{It is also worth emphasizing a
  significant difference between constrained and unconstrained
  condensation.  In the unconstrained case, a few variables carry a
  finite fraction of the sum, while in the constrained case, only a
  single variable takes a macroscopic fraction of the sum (note that
  the situation may be different, though, in the presence of
  correlations between the variables \cite{EHM06}).  We thus see that
  the 'condensation' phenomenon we define here as a non-vanishing
  value of the participation ratio in the infinite $N$ limit is a weak
  notion of condensation, which is more general than the standard
  condensation reported in the constrained case. In particular, this
  weak condensation effect does not imply the existence of a proper
  condensate, that is a 'bump' in the tail of the marginal
  distribution $p(m)$ with a vanishing relative width. The bump may
  have a non-vanishing relative width, or may even not exist, the
  distribution $p(m)$ being monotonously decreasing in this case (see
  \cite{EMZ06} for an exactly solvable example).  When relevant, we
  shall emphasize this specific character of the condensation by using
  the term 'weak condensation'.}

The goal of the present work is to understand the relation between
these two cases, which differ only by the presence or absence of a
constraint on the total mass, but yield opposite ranges of values of
$\alpha$ for the existence of condensation. To better grasp the nature
of this difference, we study here the case where the total mass is
fixed to a superextensive value $M \sim N^{1+\delta}$, with
$\delta>0$, \eb{thus extending some of the results presented in \cite{EMZ06}}. The choice of a superextensive mass is motivated by the
fact that in the unconstrained case, the total mass $M=\sum_{i=1}^N
m_i$, being the sum of iid broadly distributed variables, typically
scales superextensively, as $N^{1/\alpha}$, for $\alpha <1$. This
suggests that the case of a superextensive fixed mass may be closer to
the unconstrained case, and that the value $1+\delta=1/\alpha$ may
play a specific role. This will be confirmed by the detailed
calculations presented in Sec.~\ref{superextensive}.  Yet, before
dealing with the superextensive mass case, we will first recall in
Sec.~\ref{unconstrained} how to compute the participation ratios in
the case of unconstrained condensation, and present in
Sec.~\ref{constrained} a simplified evaluation of the participation
ratio in the case of constrained condensation with an extensive fixed
mass.


\section{Unconstrained condensation}
\label{unconstrained}

In the unconstrained case, where the masses are simply independent and
identically distributed random variables with a broad distribution,
the evaluation of the average participation ratio $Y_k$ is well-know
and has been performed using different methods
\cite{Bouchaud97,Derrida97}.  We sketch here the derivation of $Y_k$
using the auxiliary integral method put forward in \cite{Derrida97}.
Noting that
\be
\frac{1}{\left( \sum_{i=1}^N m_i\right)^k} =
\frac{1}{\Gamma(k)} \int_0^{\infty} dt \, t^{k-1} \, \exp\left(-t \sum_{i=1}^N m_i\right),
\ee
one obtains, using the property that the random variables
$m_i$ are independent and identically distributed,
\be \label{eq:Yk:unconstrained0}
Y_k = \frac{N}{\Gamma(k)} \int_0^{\infty} dt \, t^{k-1}
\langle e^{-t m} \rangle_{\alpha}^{N-1} \, \langle m^k e^{-tm} \rangle_{\alpha}
\ee
where the brackets $\langle \dots \rangle_{\alpha}$ indicate an
average over a single variable $m$ with distribution given in
Eq.~(\ref{eq:local-dist}); $\Gamma$ is the Euler Gamma function,
defined as $\Gamma(k)=\int_0^{\infty} dt \, t^{k-1} e^{-t}$.  For
large $N$, the factor $\langle e^{-t m} \rangle_{\alpha}^{N-1}$ in
Eq.~(\ref{eq:Yk:unconstrained0}) takes very small values except if $t$
is small, in which case $\langle e^{-t m} \rangle_{\alpha}$ is close
to $1$. Using a simple change of variable, one finds for $0<\alpha<1$
\cite{Derrida97}
\be
1-\langle e^{-t m} \rangle_{\alpha} \approx a \, t^{\alpha}  \qquad (t \to 0)
\ee
with $a=A\Gamma(1-\alpha)/\alpha$,
so that for $N \to \infty$,
\be
\langle e^{-t m} \rangle_{\alpha}^{N-1} \approx e^{-Nat^{\alpha}},
\ee
again for small $t$. In a similar way, one also obtains for $k > \alpha$ \cite{Derrida97}
\be
\langle m^k e^{-tm} \rangle_{\alpha} \approx A \Gamma(k-\alpha) \, t^{-(k-\alpha)}  \qquad (t \to 0).
\ee
One thus has for large $N$
\be
Y_k \approx \frac{N A \Gamma(k-\alpha)}{\Gamma(k)}
\int_0^{\infty} dt \, t^{\alpha-1} \, e^{-Nat^{\alpha}}
\ee
Using now the change of variable $v=Nat^{\alpha}$, the last integral can
be expressed in terms of the Gamma function, eventually leading, in
the limit $N \to \infty$, to \cite{Bouchaud97,Derrida97},
\be \label{eq:Yk:unconstrained}
Y_k = \frac{\Gamma(k-\alpha)}{\Gamma(k) \Gamma(1-\alpha)}
\qquad (0<\alpha<1).
\ee
The participation ratio $Y_k$ is thus non-zero for $0<\alpha<1$, and
goes to zero linearly when $\alpha \to 1$.  A similar calculation in
the case $\alpha>1$ yields $Y_k =0$ in the limit $N \to \infty$. Hence
condensation occurs for $0<\alpha<1$ in the unconstrained case. As we
shall see below, the opposite situation occurs in the constrained
case.


\section{Constrained condensation}
\label{constrained}

We now turn to the computation of the participation ratio $Y_k$ when
the total mass in the system is constrained to have the extensive
value $M=\rho N$, as a function of the exponent $\alpha$ and of the
density $\rho$.  Evaluating $Y_k$ as defined in
Eq.~(\ref{eq:part-ratios}) by averaging over the constrained
probability distribution given in Eq.~(\ref{eq:constrained-prob}), the
denominator is a constant and can be factored out of the average,
yielding the simple result:
\be 
Y_k = \frac{1}{\rho^k N^{k-1}}\langle m^k \rangle= \frac{1}{\rho^k N^{k-1}} \int_0^{\infty} dm \, p(m) \, m^k
\label{eq:y2_def}
\ee
where the marginal distribution $p(m)$ is defined as 
\be 
p(m) = f_\alpha(m) \, \frac{\mZ_{N-1}\left(\rho - \frac{m}{N}\right)}{\mZ_N(\rho)}. 
\label{eq:margina-rhom}
\ee 
Before discussing what happens for the range of exponents $\alpha$
where \eb{constrained} condensation takes place, let us briefly explain why for
$\alpha<1$ the presence of the constraint removes the
condensation and the participation ratio in Eq.~(\ref{eq:y2_def})
vanish when $N \to \infty$.

\subsection{$\alpha<1$: Absence of condensation}

The first important issue to clarify is why the condensation
  taking place in the unconstrained case for values of the power-law
  exponent $\alpha$ [see Eq.~(\ref{eq:local-dist})] in the range
  $0<\alpha<1$, then disappear when a constraint on the total mass
  value is applied. Why the constraint forces the system to stay in
  the homogeneous phase? To answer this question, it is useful to recall
the expression of the partition function of the model in terms of its
inverse Laplace transform:
\be
\mZ_N(\rho) = \frac{1}{2\pi
  i}\int_{s_0-i\infty}^{s_0+i\infty} ds \, \exp\lbrace N[ \log g_\alpha(s)+\rho
s]\rbrace,
\label{eq:inverse-Laplace}
\ee
where
\be \label{eq:def:g}
g_\alpha(s)=\int_0^\infty dm f_\alpha(m) e^{-sm}.
\ee
The values of $\rho$ in
the homogeneous ``fluid'' phase are those for which the integral in
Eq.~(\ref{eq:inverse-Laplace}) can be solved with the saddle-point
method.
In contrast, \eb{constrained} condensation occurs for all values of $\rho$ such that the saddle-point equation
\be 
\rho = -\frac{g_\alpha'(s)}{g_\alpha(s)} = \frac{\int_0^\infty
  dm\, f_\alpha(m) \, m \, e^{-sm}}{\int_0^\infty dm \, f_\alpha(m) \, e^{-sm}}
= \frac{\langle m \, e^{-sm}\rangle_\alpha}{\langle e^{-sm}\rangle_\alpha},
\label{eq:saddle-point}
\ee
admits no solution on the real axis.  It can be checked by inspection
that the function $h(s)=\log g_\alpha(s)+\rho s$ has a branch cut in
the complex $s$ plane coinciding with the negative part of the real
axis. The domain over which $s$ can be varied to look for a solution
of the saddle point equation is the positive semiaxis $[0,+\infty[$.
    When increasing $s$, the function $\langle
    m\,e^{-sm}\rangle_\alpha / \langle e^{-sm} \rangle_\alpha$
    monotonically decreases from its value $\langle m
    \rangle_{\alpha}$ reached for $s \to 0$ to $0$ for $s \to \infty$.

At this point we just need to recall that for $0<\alpha<1$ one has $\langle
m \rangle_\alpha =\infty$. This means that it is possible to
find a value $s^*$ which is a solution of Eq.~(\ref{eq:saddle-point})
for any given value of $\rho$. Hence the integral representation of
the partition function in Eq.~(\ref{eq:inverse-Laplace}) can always be
treated in the saddle-point approximation, so that condensation, which is
related to the breaking of the saddle-point approximation, never
occurs. By exploiting the saddle-point approximation for the partition
function it is then not difficult to compute explicitly the expression
in Eq.~(\ref{eq:margina-rhom}), which reads
\be 
p(m) \sim \frac{e^{-s^*m}}{m^{1+\alpha}},
\label{eq:margina-rhom-A}
\ee
where $s^*<\infty$ is the solution of the saddle point
equation. The vanishing of the participation ratio $Y_k$
  then follows easily, according to its expression in
  Eq.~(\ref{eq:y2_def}), from the presence of the exponential cutoff
  in $p(m)$ [Eq.~(\ref{eq:margina-rhom-A})]:
\be
Y_k  = \frac{\langle m^k \rangle}{\rho^k N^{k-1}},
\ee
where $\langle m^k \rangle = \int_0^{\infty}dm \,m^k \,p(m)$ is a function 
which does not depend on $N$ in the large $N$ limit, so that 
\be 
\lim_{N\rightarrow\infty} \frac{\langle m^k \rangle}{\rho^k N^{k-1}}=0.
\ee
We have therefore seen that for any $\alpha<1$, if one
constrains the system to have an extensive mass $M=\rho N$, the participation ratio vanishes in the thermodynamic limit: $\lim_{N\rightarrow\infty} Y_k = 0$,
as expected since no condensation occurs in this case.

\subsection{$\alpha > 1$: Homogeneous phase at low density ($\rho<\rho_c$)}

As soon as the exponent $\alpha$ of the single-variable distribution
is increased above $\alpha_c=1$, namely as soon as the first moment of
the distribution $f_\alpha(m)$ becomes finite, a condensed
phase appears at finite $\rho_c$. For any given value of $\alpha$
the critical density $\rho_c$ is the maximal density for which the
saddle point equation Eq.~(\ref{eq:saddle-point}) has a solution.  As
we have already noticed in the previous section, the maximum value
which can be attained by the term on the right of
Eq.~(\ref{eq:saddle-point}) is $\langle m \rangle_\alpha$, so that for
all values $\rho>\rho_c=\langle m \rangle_\alpha$ the saddle-point
approximation breaks down and one has condensation~\cite{MEZ05,EMZ06}.
Nevertheless, for $\alpha>1$ and $\rho<\rho_c$ the system is
still in the homogeneous phase and, similarly to what is done in the
case $\alpha<1$, one can compute the marginal distribution
$\rho(m)$ according to its definition in Eq. (\ref{eq:margina-rhom})
by using the saddle-point approximation.  The result is also in this
case
\be 
p(m) \sim f_\alpha(m)\, e^{-m/m_0},
\label{eq:rhom-fluid}
\ee
and we know from~\cite{MEZ05,EMZ06} that the characteristic mass $m_0$
diverges when the density $\rho$ tends to $\rho_c$ as $m_0 \sim (\rho-\rho_c)^{-1}$ for
$\alpha>2$ and as $m_0 \sim (\rho-\rho_c)^{-1/(\alpha-1)}$ for $1
< \alpha < 2$. \\

\begin{figure}
\includegraphics[width=0.85\columnwidth]{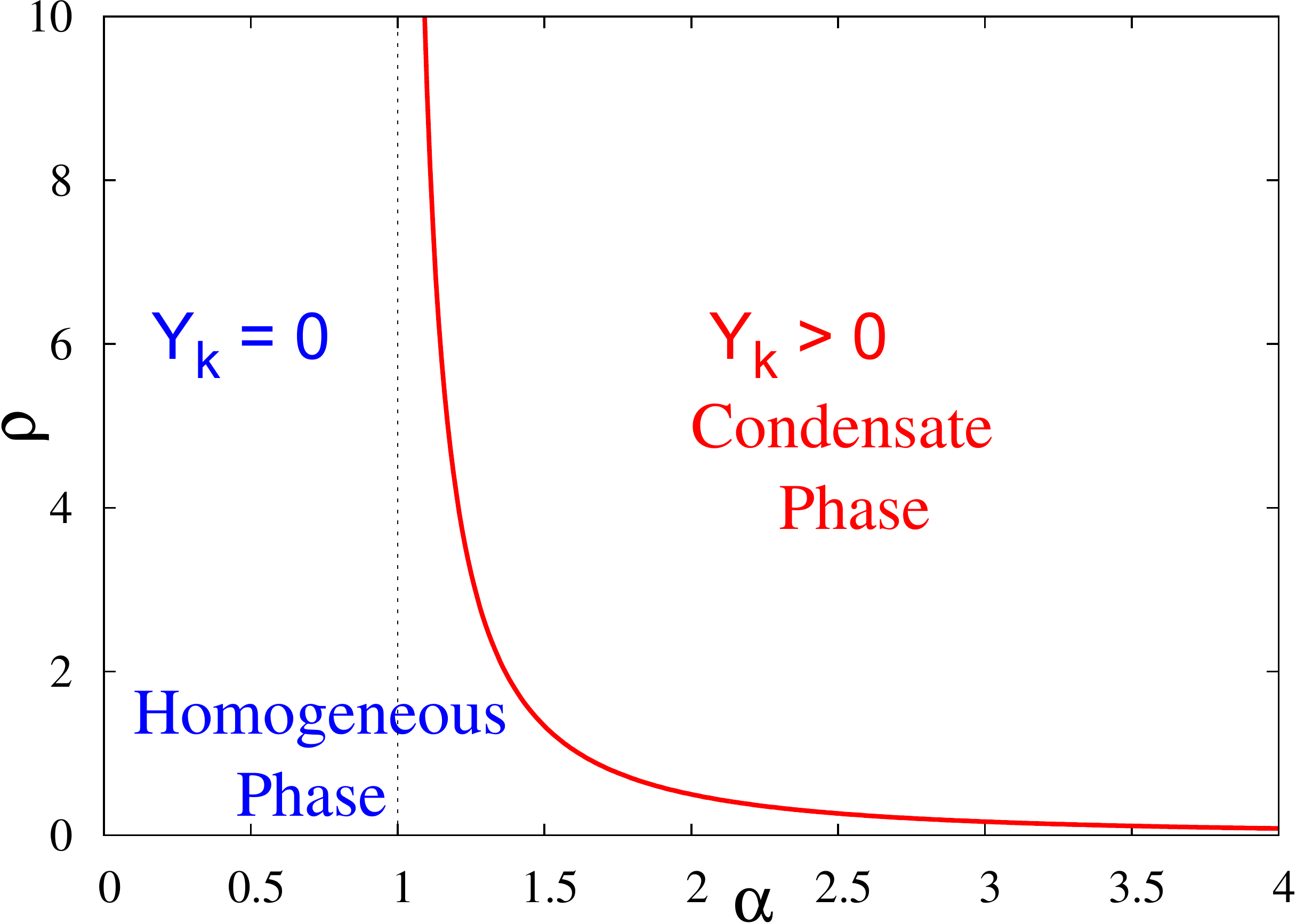}
\caption{
Phase diagram for the values of the participation rations in
  the ($\alpha$, $\rho$) plane in the presence of an extensive constraint on the total value of the mass $\sum_{i=1}^N m_i = \rho N$. The (red)
  continuous line marks the separation of the condensed phase
  ($\lim_{N \to \infty} Y_k > 0$) and the homogeneous phase
  ($\lim_{N \to \infty} Y_k = 0$). The vertical dotted (black) line marks the critical value
  $\alpha_c=1$ where the critical density $\rho_c$ for condensation diverges.}
\label{fig0}
\end{figure}

\subsection{$\alpha> 1$: Condensed phase at high density ($\rho>\rho_c$)}

Let us now study what happens in the condensed phase.  For $\alpha> 1$
and $\rho>\rho_c$, one observes for large $N$ a coexistence between a
homogeneous fluid phase carrying a total mass approximately equal to
$N\rho_c$ and a condensate of mass $M_{\textrm{cond}} \approx
(\rho-\rho_c) N$.  The marginal distribution $p(m)$ can be
approximately written as \cite{EM08}
\be \label{eq:pm:condensate}
p(m) \approx f(m) + p_{\textrm{cond}}(m,\rho,N)
\ee
where $p_{\textrm{cond}}(m,\rho,N)$ is the mass distribution of the
condensate, normalized according to $\int_0^{\infty}
p_{\textrm{cond}}(m,\rho,N) = 1/N$ to account for the fact that the
condensate is present on a single site.  It has been shown
\cite{MEZ05,EMZ06} that for $\alpha>2$, the distribution
$p_{\textrm{cond}}(m,\rho,N)$ is Gaussian, with a width proportional
to $N^{1/2}$. In other words, the condensate exhibits normal
fluctuations.  In contrast, for $1<\alpha<2$, the distribution
$p_{\textrm{cond}}(m,\rho,N)$ has a broader, non-Gaussian shape, with
a typical scale of fluctuation $\sim N^{1/\alpha}$ \cite{MEZ05,EMZ06}.
For all values of $\alpha >1$, however, the relative fluctuations of
$M_{\textrm{cond}}$ vanish in the large $N$ limit:
\bea \nonumber
\frac{M_{\textrm{cond}} - \overline{M}_{\textrm{cond}}}{M_{\textrm{cond}}} &\sim& N^{-(\alpha-1)/\alpha} \qquad \, \textrm{if} \quad 1<\alpha<2\\
&\sim& N^{-1/2} \qquad \qquad \textrm{if} \quad \alpha >2
\eea
with $\overline{M}_{\textrm{cond}} = (\rho-\rho_c) N$.
Hence in order to compute the large $N$ behavior of moments of the
distribution $p(m)$, one can further approximate $p(m)$ as
\be \label{eq:pm:condensate2}
p(m) \approx f(m) + \frac{1}{N}\, \delta\left( \sum_{i=1}^N m_i - \overline{M}_{\textrm{cond}} \right)
\ee
Note that more accurate expressions of the distribution $p(m)$ can be found in \cite{EMZ06}.

From Eq.~(\ref{eq:pm:condensate2}), the moment $\langle m^k \rangle$ is evaluated as
\be \label{eq:mk:condensate}
\langle m^k \rangle \approx \int_0^{\rho_c N} dm \, m^k f(m) + \frac{\overline{M}_{\textrm{cond}}^k}{N} .
\ee
The integral in Eq.~(\ref{eq:mk:condensate}), corresponding to the
fluid phase contribution to the moment, has a different scaling with
$N$ depending on the respective values of $k$ and $\alpha$. If $k
<\alpha$, the integral converges to a finite limit when $N$ goes to
infinity. On the contrary, when $k >\alpha$, the integral diverges
with $N$ and scales as $N^{k-\alpha}$.

The participation ratio reads $Y_k=\langle m^k \rangle/(\rho^k
N^{k-1})$, so that the contribution of the fluid phase to the
participation ratio scales as $1/N^{k-1}$ for $k <\alpha$, and as
$1/N^{\alpha-1}$ for $k >\alpha$; in both cases, this contribution
vanishes for $N \to \infty$, when $k>1$ and $\alpha >1$.  The
remaining contribution, resulting from the condensate, simply leads to
\be \label{eq:Yk:constrained:extensive}
Y_k = \left( \frac{\rho-\rho_c}{\rho} \right)^k ,
\ee
where we recall that $\rho>\rho_c$ in the condensed phase.  Hence
$Y_k$ goes to zero at the onset of condensation ($\rho \to \rho_c$),
so that the transition can be thought as continuous if one considers
the participation ratio as an order parameter. In the opposite limit
$\rho \to \infty$, the participation ratio goes to $1$, indicating a
full condensation.

A phase diagram in the $(\alpha,\rho)$-plane summarizing the results
of this section for the case of constrained condensation with
extensive mass is shown in Fig.~\ref{fig0}.

 
\section{Constraint to a superextensive total mass $M=\tilde{\rho} N^{1+\delta}$}
\label{superextensive}

As explained in the introduction, the unconstrained condensation
occurs for $\alpha <1$, while the contraint-driven condensation occurs
at $\alpha >1$ (and at high enough density). Given that the typical
total mass in the unconstrained case is superextensive for $\alpha
<1$, it is of interest to study condensation effects in the more
general case of a fixed superextensive total mass $M=\tilde{\rho}
N^{1+\delta}$, with $\delta>0$ and $\tilde{\rho}$ a parameter which
generalizes the usual notion of density.  The joint probability
distribution $p(m_1,\ldots,m_N|\tilde{\rho})$ reads in this case
\be 
p(m_1,\ldots,m_N|\tilde{\rho}) = \frac{1}{\mZ_{N,\delta}(\tilde{\rho})} \prod_{i=1}^N f(m_i)
\delta\left[ \sum_{i=1}^N m_i - \tilde{\rho} N^{1+\delta} \right],
\label{eq:constrained-prob-super}
\ee
where $\mZ_{N,\delta}(\tilde{\rho})$ is a normalization
  factor [see Eq.~(\ref{eq:canonical-partition})].
We wish to determine for which values of $\alpha$ and
$\tilde{\rho}$ condensation occurs in this case, using as an order parameter for condensation the participation ratio $Y_k = N\langle m^k \rangle/M^k$, which reads in the present case as
\be 
Y_k = \frac{\langle m^k \rangle}{\tilde{\rho}^k N^{k-1+k\delta}}= \frac{1}{\rho^k N^{k-1+k\delta}} \int_0^{\infty} dm \, p(m) \, m^k.
\label{eq:participation-super}
\ee
The expression of the marginal distribution $p(m)$ in the case of the
superextensive total mass is given below in Eq.~(\ref{eq:pm-super}).

In the following, we first use in Sec.~\ref{sec:crit:line:super} the integral representation of the partition function $\mZ_N(\tilde{\rho})$ 
\eb{in order to get indications on the phase diagram}
in the $(\alpha,\delta)$ plane.
\eb{This preliminary analysis will suggest the existence of a transition line, that will be confirmed in Sec.~\ref{super-extensive:homogeneous} to \ref{super-extensive:critical} by an explicit determination of the marginal distribution $p(m)$ and the participation ratio $Y_k$ respectively below, above and on the anticipated transition line.}

\eb{
\subsection{Preliminary analysis of the phase diagram} }
\label{sec:crit:line:super}

We start by expressing the partition function $\mZ_{N,\delta}(\tilde{\rho})$ as an integral representation in terms of its inverse Laplace transform.
The Laplace
transform $\hat{\mZ}_{N,\delta}(s)$ of $\mZ_{N,\delta}(\tilde{\rho})$
is expressed as \be \label{eq:ZNs:superext} \hat{\mZ}_{N,\delta}(s)
\equiv \int_0^{\infty} d\tilde{\rho}\, e^{-s\tilde{\rho}}
\mZ_{N,\delta}(\tilde{\rho}) = \frac{1}{N^{1+\delta}} \,
g_\alpha\left(\frac{s}{N^{1+\delta}} \right)^N \ee where $g_\alpha(s)$
is defined in Eq.~(\ref{eq:def:g}).  After a simple change of
variable, the inverse Laplace representation of the partition function
$\mZ_N(\tilde{\rho})$ reads
\be 
\mZ_{N,\delta}(\tilde{\rho}) = \frac{1}{2\pi i}\int_{s_0-i\infty}^{s_0+i\infty} ds \, \exp\lbrace N[ \log g_\alpha(s)+\tilde{\rho} N^\delta s]\rbrace.
\label{eq:inverse-Laplace-super}
\ee
The value of $s_0$, although arbitrary, can be conveniently chosen to
be the saddle-point value of the argument of the exponential in
Eq.~(\ref{eq:inverse-Laplace-super}), when a saddle-point $s_0>0$
exists (this is due to the presence of a branch-cut singularity on the
negative real axis, as discussed in the previous section).  When no
saddle-point exists, the equivalence between canonical and
grand-canonical ensemble breaks down, and condensation is expected to
occur.  A saddle-point of the integral in
Eq.~(\ref{eq:inverse-Laplace-super}) should satisfy the following
equation,
\be 
\tilde{\rho} N^\delta = -\frac{g_\alpha'(s)}{g_\alpha(s)} = \frac{\langle
  m \, e^{-sm} \rangle_\alpha}{\langle e^{-sm}\rangle_\alpha}.
\label{eq:saddle-point-super}
\ee
Note that this approach is heuristic, since the saddle-point
should in principle not depend on $N$. However, the $N$-dependence is
not a problem when testing the existence of a saddle-point. If it
exists, the saddle-point evaluation of the integral then requires a
change of variable for (some power of) $N$ to appear only as a global
prefactor in the argument of the exponential.

For $\alpha>1$, we know from the results of section~\ref{constrained}
that the saddle-point equation (\ref{eq:saddle-point-super}) has a
solution only if $\tilde{\rho} N^\delta < \rho_c$. This condition is
never satisfied for $\delta >0$ and $N \to \infty$, so that no
saddle-point exists and condensation occurs for any value of
$\tilde{\rho}>0$ when $\alpha>1$.

The situation is thus quite similar to the extensive mass case: there
is a homogeneous phase carrying a total mass $N \rho_c$ which coexists
with a superextensive condensate with a mass $M_c = \tilde{\rho}
N^{1+\delta} - N \rho_c \approx \tilde{\rho} N^{1+\delta}$, so that
the condensate carries a fraction of the total mass equal to one in
the limit $N \to \infty$. It follows that the participation ratio $Y_k
=1$ in this limit.

In contrast, for $\alpha <1$, the function $g_\alpha'(s)/g_\alpha(s)$
spans the whole positive real axis, and the saddle-point equation
(\ref{eq:saddle-point-super}) always has a solution $s_0(N)$, which
goes to $0$ when $N \to \infty$.  One then has to factor out
the $N$-dependence through an appropriate change of variable, and to
check whether a saddle-point evaluation of the integral can be
made. For $s \to 0$, one has $g_\alpha'(s)/g_\alpha(s) \approx a
\alpha s^{\alpha-1}$, so that $s_0 \sim N^{-\delta/(1-\alpha)}$.
Using the change of variable $s=z N^{-\delta/(1-\alpha)}$ in the
integral appearing in Eq.~(\ref{eq:saddle-point-super}), the argument
of the exponential can be rewritten as
\be
N^{1-\alpha\delta/(1-\alpha)} (-az^{\alpha}+\tilde{\rho} z), \quad 
\ee
with $a=A \,\Gamma(1-\alpha)/\alpha$, and where we have used the
small-$s$ expansion $g_{\alpha}(s) \approx 1-as^{\alpha}$, valid for
$0<\alpha<1$.  The saddle-point evaluation of the integral is valid
only if the $N$-dependent prefactor diverges, meaning that
$1-\alpha\delta/(1-\alpha)>0$, or equivalently
\be \label{def-deltac}
\delta < \delta_c = \frac{1-\alpha}{\alpha}.
\ee
Hence for $\alpha <1$ and $\delta < \delta_c$, a saddle-point
evaluation of the partition function is possible, and the equivalence
between canonical and grand-canonical ensembles holds: the system is
in the homogeneous phase, and no condensation occurs. 
For $\delta \ge \delta_c$, \eb{the saddle-point evaluation of the partition function is no longer possible, which suggests that the equivalence of ensembles breaks down. This is an indication that condensation may occur.
We show through explicit calculations in Sec.~\ref{super-extensive:condensed} and Sec.~\ref{super-extensive:critical} that condensation occurs when $\delta \ge \delta_c$, in the sense that the participation ratio $Y_k$ takes a nonzero value in the infinite $N$ limit.}

\begin{figure}
\includegraphics[width=0.85\columnwidth]{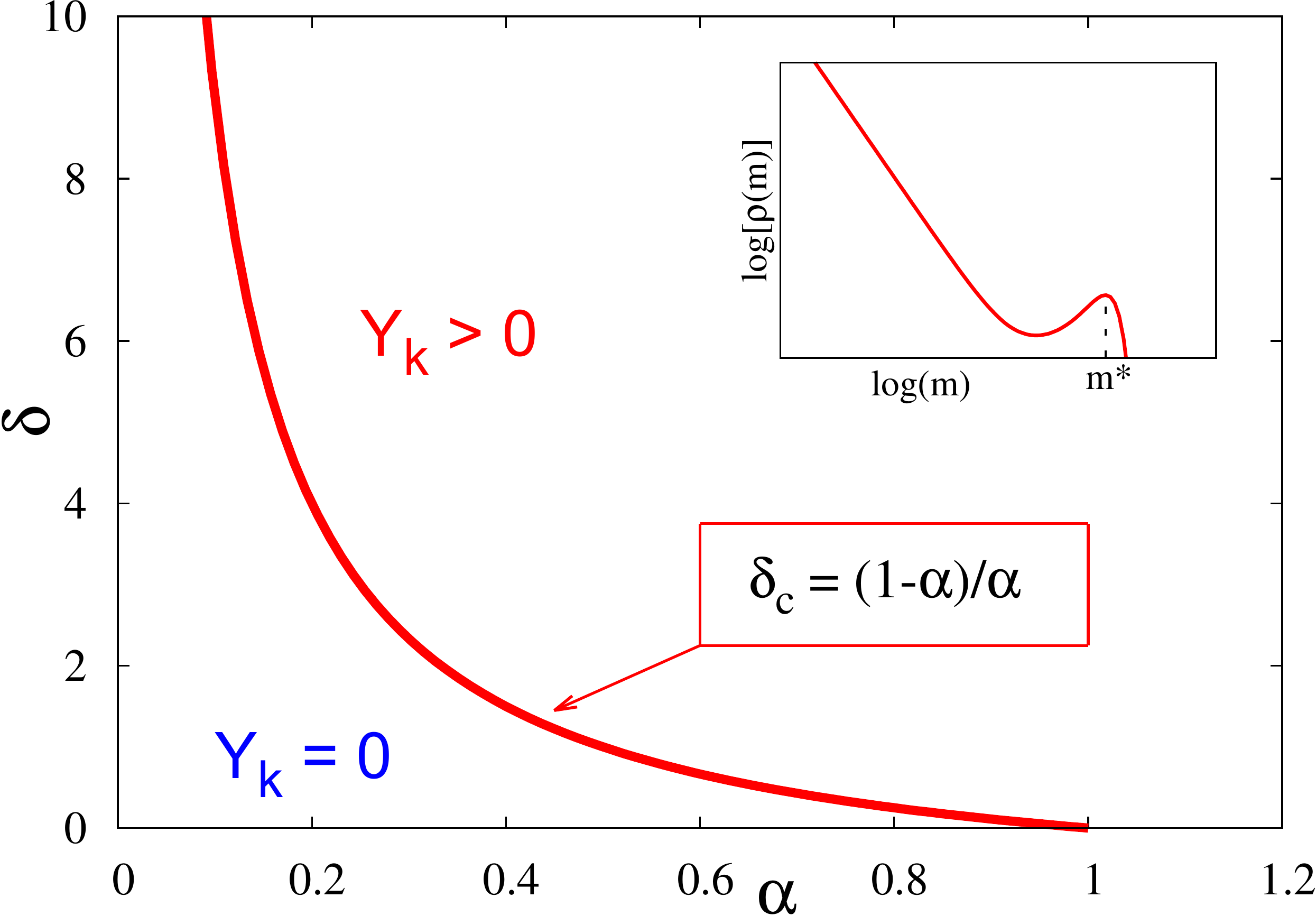}
\caption{\emph{Main}: Phase diagram for the values of the
  participation ratio in the ($\alpha$, $\delta$) plane in the
  presence of a super-extensive constraint on the total value of
  the mass $\sum_{i=1}^N m_i = \tilde{\rho} N^{1+\delta}$. The (red)
  continuous line marks the separation of the condensed phase
  ($\lim_{N \to \infty} Y_k > 0$) and the homogeneous phase ($\lim_{N
    \to \infty} Y_k = 0$). \emph{Inset}: schematic representation of
  the marginal probability distribution of the local mass, $p(m)$, in
  the presence of condensation, namely in the whole region $\delta
  \geq \delta_c$.}
\label{fig1}
\end{figure}

Before proceeding to a
detailed characterization of this condensation, let us briefly comment
on the value of $\delta_c$.  For $\delta=\delta_c$, the total mass in
the system scales as $M \sim N^{1/\alpha}$, and this scaling precisely
corresponds to the typical value of the total mass present in the
unconstrained case (see section~\ref{unconstrained}), \eb{as already noticed in \cite{EMZ06}}. Hence $\delta <
\delta_c$ corresponds to imposing a total mass much smaller than the
'natural' unconstrained mass, while for $\delta > \delta_c$ one
imposes a mass much larger than the typical unconstrained mass,
leading to condensation. In this sense, the situation is similar to
that of the extensive mass case for $\alpha >1$, where condensation
occurs when a mass $M$ larger that the unconstrained mass $N\rho_c$ is
imposed. In Fig.~\ref{fig1} we present the phase diagram of
  the model for the case of a constraint to a superextensive total mass,
  which is a phase diagram in the plane $(\alpha,\delta)$. Two
  observations are in order for this phase diagram. First, as will be
  explained in detail in in Sec.~\ref{super-extensive:condensed} and
  Sec.~\ref{super-extensive:critical}, the presence of a condensed
  phase never depends on the value of the parameter
  $\tilde{\rho}$. 
Second, a remarkable difference with the case of
constrained condensation with extensive mass is that on the \eb{transition} line $\delta_c(\alpha)$ (see Fig.~\ref{fig1}), the system is in the condensed phase, as will be explained in Sec.~\ref{super-extensive:critical}.
This behavior is in contrast with the critical line $\rho_c(\alpha)$ corresponding to an extensive mass (see the phase diagram in Fig.~\ref{fig0}), along which the system is \emph{not} in the condensed phase.

Below, we evaluate the distribution $p(m)$ and the participation ratio
$Y_k$ for $\alpha <1$ in the three cases $\delta < \delta_c$, $\delta
> \delta_c$ and $\delta = \delta_c$ respectively.\\

\subsection{Case $\delta < \delta_c$: Homogeneous phase}
\label{super-extensive:homogeneous}

As we have seen above, the system remains homogeneous for $\alpha <1$ and $\delta < \delta_c$, and the equivalence between canonical and grand-canonical ensembles holds.
One can thus more conveniently perform calculations in the grand-canonical ensemble, with a chemical potential $\mu_N$ which depends on $N$, and which will be determined below as a function of the total mass.
The single-mass distribution $p(m)$ simply reads
\be
p(m) \approx f(m)\, e^{-\mu_N m}  \qquad (0<m<M),
\ee
where we have neglected the correction to the normalization factor, as the latter remains very close to $1$ since $\mu_N$ is very small.
Note that the distribution $p(m)$ monotonously decays at large $m$ as
\be
p(m) \approx \frac{A}{m^{1+\alpha}} \, e^{-\mu_N m}
\ee
as is typical for a homogeneous phase.
The $k$-th moment of this distribution is obtained for $k>\alpha$ as
\bea
\langle m^k \rangle &=& \int_0^M dm \, m^k p(m) \\ \nonumber
&\approx& A \int_0^{\infty} dm \, m^{k-1-\alpha} e^{-\mu_N m}
= \frac{A \, \Gamma(k-\alpha)}{\mu_N^{k-\alpha}} .
\eea
The value of $\mu_N$ is then determined from the condition
$\langle m \rangle = M/N = \tilde{\rho} N^{\delta}$, yielding
\be
\mu_N = \left[ \frac{A}{\tilde{\rho}} \, \Gamma(1-\alpha)\right]^{1/(1-\alpha)} \, N^{-\delta/(1-\alpha)} .
\ee
The participation ratio $Y_k$ is then given by
\be
Y_k = K \tilde{\rho}^{(k-1)/\delta_c} \, N^{-(k-1)(\delta_c-\delta)/\delta_c}
\ee
with
\be
K = A^{1-\omega} \, \frac{\Gamma(k-\alpha)}{\Gamma(1-\alpha)^\omega}, \qquad \omega = \frac{k-\alpha}{1-\alpha},
\ee
and where $\delta_c$ is defined in Eq.~(\ref{def-deltac}). One thus
obtains that $Y_1=1$ as it should, and that for $k>1$, $Y_k \to 0$
when $N \to \infty$, which confirms the absence of condensation for
$\delta < \delta_c$. Yet, it is interesting to note that the decay of
$Y_k$ becomes slower when increasing $\delta$, and becomes
approximately logarithmic in $N$ when $(\delta_c-\delta)/\delta_c \ll
1$.

\subsection{Case $\delta > \delta_c$: Condensed phase}
\label{super-extensive:condensed}

When $\delta > \delta_c$ (and $\alpha <1$), the partition function
$Z_{N,\delta}(\tilde{\rho})$ can no longer be evaluated by a
saddle-point method, and equivalence of ensembles breaks down, so that
one has to work in the canonical ensemble.  From
Eq.~(\ref{eq:ZNs:superext}), the Laplace transform
$\hat{\mZ}_{N,\delta}(s)$ reads in the large $N$ limit, using the
small-$s$ behavior $g_\alpha(s) \approx 1-as^\alpha$,
\be \label{eq:ZNs:superext:supercrit}
\hat{\mZ}_{N,\delta}(s) = N^{-1-\delta} \, \left( 1 - \frac{a s^{\alpha}}{N^{\nu}} \right)
\ee
with $\nu=\alpha (1+\delta)-1$ (note that
Eq.~(\ref{eq:ZNs:superext:supercrit}) is not restricted to small $s$
values). 

By assuming a scaling function $G(x)$ which satisfies the
  normalization condition $\int_0^{\infty} G(x) \, dx =1$ and which
  has the asymptotic behaviour $G(x) \approx A/x^{1+\alpha}$ for $x
  \to \infty$, one can try to write the partition function
  $Z_{N,\delta}(\tilde{\rho})$ in direct space as
\be \label{eq:ZN:superext:supercrit}
Z_{N,\delta}(\tilde{\rho}) \approx N^{-1-\delta+\nu/\alpha} G(\tilde{\rho} N^{\nu/\alpha}).
\ee
It is then not difficult to check that the expression in
Eq.~(\ref{eq:ZN:superext:supercrit}) is (asymptotically in $N$) the
correct one: the expansion for small $s$ of its Laplace transform
corresponds precisely to the expression of $\hat{\mZ}_{N,\delta}(s)$
in Eq.~(\ref{eq:ZNs:superext:supercrit}).

From the knowledge of $Z_{N,\delta}(\tilde{\rho})$ one can then compute the distribution $p(m)$, which reads as
\bea
p(m) &=& \frac{f(m)}{Z_{N,\delta}(\tilde{\rho})} \, Z_{N-1,\delta}\left( \tilde{\rho}-\frac{m}{(N-1)^{1+\delta}} \right) \nonumber \\
&\approx& \frac{f(m)}{G(\tilde{\rho} N^{\delta-\delta_c})} \, G\left( \tilde{\rho} N^{\delta-\delta_c} - \frac{m}{N^{1/\alpha}} \right)
\label{eq:pm-super}
\eea
and has a non-monotonous shape, as seen by evaluating $p(m)$ in the regime
$m \sim xM$ with $0<x<1$, which leads to
\be
p(xM) \approx \frac{A}{N^{(1+\delta)(1+\alpha)} \tilde{\rho}^{1+\alpha} [x(1-x)]^{1+\alpha}}.
\ee
Note that the divergences at $x=0$ and $x=1$ are regularized for
values of $x$ such that $x \sim M^{-1}$ and $1-x \sim M^{-1}$
respectively.  The non-monotonic shape of $p(xM)$, which is
schematically represented in the inset of Fig.~\ref{fig1}, is a strong
similarity that the constrained condensation for $\alpha < 1$ (and
superextensive total mass) bears with the constrained condensation for
$\alpha>1$ (and extensive total mass). At the same time such a
non-monotonic shape of $p(m)$ is a remarkable qualitative difference
with the case of unconstrained condensation found for the same range
of the exponent, $\alpha<1$, in which case the local mass distribution
decays monotonously at large values as $p(m) \sim 1/m^{1+\alpha}$.
\eb{Interestingly, the expression (\ref{eq:pm-super}) of $p(m)$ can be
  rewritten as 
\be p(m) \approx \frac{f(m)}{G(\tilde{\rho}
    N^{\delta-\delta_c})} \, G\left(\frac{M-m}{N^{1/\alpha}} \right)
\ee
with $M=\tilde{\rho} N^{1+\delta}$, which shows that the `bump'
  occuring for $m \approx M$ has a width $\sim N^{1/\alpha}$. Hence
  its relative width scales as $N^{1+\delta}/N^{1/\alpha} =
  1/N^{\delta-\delta_c}$ and thus goes to zero when $N \to \infty$ for
  $\delta >\delta_c$.  It would thus be legitimate in this case to
  call the bump a condensate, because it has a well-defined mass
  $M_{\rm cond} \sim M$.

To complete the analysis, let us compute the participation ratio $Y_k$.}
In the large $N$ limit, the moment $\langle m^k \rangle = \int_0^M dm \, m^k p(m)$ can be computed as,
using the change of variable $v= \tilde{\rho} N^{\delta-\delta_c} - m/N^{1/\alpha}$,
\be \label{eq:mk:superext:supercrit}
\langle m^k \rangle \approx \frac{M^k}{N} \int_0^{M/N^{1/\alpha}} dv \,
\left( 1-\frac{vN^{1/\alpha}}{M} \right)^{k-1-\alpha} G(v)
\ee
where we have used the asymptotic (large argument) behavior of $f(m)
\sim A/m^{1+\alpha}$ and $G(x)\sim A/x^{1+\alpha}$.  It is then easy
to show (see Appendix~A) that the integral in
Eq.~(\ref{eq:mk:superext:supercrit}) tends to
$\int_0^{\infty} dv \, G(v)=1$ when $N\rightarrow \infty$.  One thus simply gets
\be \label{eq:mk:superext:supercrit2}
\langle m^k \rangle \approx \frac{M^k}{N}
\ee
which, using $Y_k = N\langle m^k \rangle/M^k$,
immediately leads to the conclusion that $Y_k=1$ in the limit $N\to \infty$.
Hence, \eb{as anticipated above}, a strong condensation occurs for $\delta > \delta_c$ and $\alpha <1$, \eb{in the sense that the condensate carries almost all the mass present in the system.}

\subsection{Case $\delta = \delta_c$: \eb{Marginal} condensed phase}
\label{super-extensive:critical}

For $\alpha <1$ and $\delta = \delta_c$, the Laplace transform of the partition function reads for large $N$
\be
\hat{\mZ}_{N,\delta_c}(s) = N^{-1/\alpha} \, e^{-as^{\alpha}}
\ee
from which the partition function is obtained as
\be
\mZ_{N,\delta_c}(\tilde{\rho}) = N^{-1/\alpha} \, H(\tilde{\rho})
\ee
where the function $H(\tilde{\rho})$ is independent of $N$ and is defined by its Laplace transform,
\be
\int_0^{\infty} d\tilde{\rho} \, e^{-s\tilde{\rho}} H(\tilde{\rho}) =  e^{-as^{\alpha}}
\ee
($H$ is actually a one-sided L\'evy distribution).
The small $s$ behavior $e^{-as^{\alpha}} \approx 1-as^{\alpha}$ implies the large $\tilde{\rho}$ behavior
\be \label{eq:asympt:Hrho}
H(\tilde{\rho}) \approx \frac{A}{\tilde{\rho}^{1+\alpha}}
\ee
where again $A$ is defined from the large $m$ behavior $f(m) \approx A/m^{1+\alpha}$.

The distribution $p(m)$ is given for large $N$ by
\be \label{eq:pm:marginal}
p(m) = \frac{f(m)}{H(\tilde{\rho})} \, H\left(\frac{M-m}{N^{1/\alpha}}\right).
\ee
with $M=\tilde{\rho} N^{1/\alpha}$.
It is interesting to evaluate $p(m)$ in the regime where $m \sim xM$ with $0<x<1$, which leads to
\be \label{eq:large:m:pm}
p(xM) \approx \frac{A}{N^{1+1/\alpha} \tilde{\rho}^{1+\alpha} H(\tilde{\rho})} \,
\frac{H\Big(\tilde{\rho}(1-x)\Big)}{x^{1+\alpha}}.
\ee
For large enough $\tilde{\rho}$, the shape of $p(m)$ is not monotonous, since
the large $\tilde{\rho}$ expansion of Eq.~(\ref{eq:large:m:pm}) yields
\be
p(xM) \approx \frac{A}{N^{1+1/\alpha} \tilde{\rho}^{1+\alpha} [x(1-x)]^{1+\alpha}}
\ee
with a regularization of the divergence appearing at $x=1$ for $1-x \sim \tilde{\rho}^{-1}$, and of the divergence at $x=0$ for $x \sim M^{-1}$.
\eb{So here again, a bump appears in the distribution, but its width scales as $N^{1/\alpha}$ as seen from Eq.~(\ref{eq:pm:marginal}), so that the relative width remains of the order of one. Following \cite{EMZ06}, one may call this bump a `pseudo-condensate'.

The above argument on the existence of the bump in the distribution $p(m)$ was based on a large $\tilde{\rho}$ limit. The explicit example studied in \cite{EMZ06} indeed shows that the bump may disappear below a certain value of $\tilde{\rho}$.}

We now turn to the evaluation of the moment $\langle m^k \rangle$.
Using the change of variable $v=\tilde{\rho} - m/N^{1/\alpha}$, as well as the asymptotic (large argument) behaviors of $f(m)$ and $H(\tilde{\rho})$, the moment $\langle m^k \rangle$ can be evaluated as
\be
\langle m^k \rangle = \tilde{\rho}^{1+\alpha} N^{-1+k/\alpha} \int_0^{\tilde{\rho}} dv \, (\tilde{\rho}-v)^{k-1-\alpha} H(v).
\ee
It follows that $Y_k = N\langle m^k \rangle/M^k$ is given by
\be \label{eq:Yk:superext:crit}
Y_k = \frac{1}{\tilde{\rho}^{k-1-\alpha}} \int_0^{\tilde{\rho}} dv \, \left(\tilde{\rho}-v\right)^{k-1-\alpha} H(v)
\ee
which is one of the main results of this paper.
Note that the convergence of the integral at the upper bound implies $k-\alpha >0$.
Note also that the integral in Eq.~(\ref{eq:Yk:superext:crit}) is a convolution, which in some cases may be conveniently evaluated using a Laplace transform, given that $H(\tilde{\rho})$ is known through its Laplace transform. For a numerical evaluation of $Y_k$, one may thus compute analytically the Laplace transform of the integral in Eq.~(\ref{eq:Yk:superext:crit}), yielding
\be \label{eq:Yk:inverse:laplace}
\mathcal{L}\Big( \tilde{\rho}^{k-1-\alpha} Y_k(\tilde{\rho})\Big)
= \frac{\Gamma(k-\alpha)}{s^{k-\alpha}} \, e^{-as^{\alpha}}
\ee
and perform numerically the inverse Laplace transform.

The Laplace transform approach is also convenient to determine analytically the small $\tilde{\rho}$ behavior of $Y_k$, since the inverse Laplace transform can be evaluated through a saddle-point calculation in this limit.
One finds
\be \label{eq:Yk:crit:small:rho}
Y_k \approx B \, \tilde{\rho}^{\lambda} \, e^{-c/\tilde{\rho}^{\alpha/(1-\alpha)}}
\ee
with parameters $\lambda$, $c$ and $B$ given by
\bea
\label{eq:def:lambda}
\lambda &=& \frac{\alpha (2k-2\alpha -1)}{2(1-\alpha)}\\
c &=& \frac{1-\alpha}{\alpha} \, (\alpha a)^{1/(1-\alpha)}\\
B &=& \frac{\Gamma(k-\alpha)}{\sqrt{2\pi(1-\alpha)}} 
\, (\alpha a)^{\frac{1+2\alpha-2k}{2-2\alpha}}
\eea
More detailed calculations on the derivation of Eq.~(\ref{eq:Yk:crit:small:rho}) are reported in Appendix~A.

In the large $\tilde{\rho}$ limit, it is easy to show that $Y_k$ goes to $1$, following a procedure similar to the one used in the case $\delta > \delta_c$. It is of interest to compute the first correction in $\tilde{\rho}$ (see Appendix~A), and one finds
\be \label{eq:Yk:crit:large:rho}
Y_k \approx 1 - \frac{B'}{\tilde{\rho}^\alpha}
\ee
with
\be
B' = \frac{A}{\alpha} \, \frac{\Gamma(1-\alpha) \, \Gamma(k-\alpha)}{\Gamma(k-2\alpha)}.
\ee
Note in particular that if $\alpha \ll 1$, the convergence of $Y_k$ to $1$ is very slow.

\eb{
In summary, one has in the case $\delta=\delta_c$ a non-standard, weak condensation effect, which does not correspond to the genuine condensation effect reported in the literature \cite{BBJ97,MKB98,GSS03,MEZ05,EMZ06,EH05,S08-leshouches,HMS09,WCBE14,EW14}. Here, the weak condensation effect simply means that the participation ratio takes a nonzero value in the infinite size limit, indicating that a few random variables carry a finite fraction of the sum. However, as mentioned above, there is no well-defined condensate that would coexist with a fluid phase. Depending on the generalized density $\tilde{\rho}$, the marginal distribution $p(m)$ either decreases monotonously, or has a bump which corresponds only to a pseudo-condensate, since the relative width of the bump remains of the order of one, see Eq.~(\ref{eq:pm:marginal}). In addition, the line $\delta=\delta_c$ does not correspond to a well-defined transition line in the $(\alpha,\delta)$ plane, in the sense that the state of the system continuously depends on the generalized density $\tilde{\rho}$, as shown by the expression of the participation ratio $Y_k$ given in Eq.~(\ref{eq:Yk:superext:crit}).}

\section{Conclusion}

The general motivation of this work was to better understand the
connection between condensation in the unconstrained case and in the
constrained case with extensive mass, because condensation occurs on
opposite ranges of the exponent $\alpha$ (which defines the power-law
decay of the unconstrained probability distribution), respectively
$\alpha<1$ and $\alpha >1$.  To this aim, we have studied condensation
in the case where the total mass is constrained to a superextensive
value $M = \tilde{\rho} N^{1+\delta}$, where $\delta > 0$, motivated
by the fact that the typical scaling of the total mass is also
superextensive, $M \sim N^{1/\alpha}$, when condensation
  takes place in the unconstrained case, which happens for
$\alpha<1$.

We indeed found that the case of a fixed superextensive total mass
interpolates in a sense between the case with a fixed extensive mass
and the unconstrained case: condensation is found for values
  of the power law exponent in the interval $0<\alpha<1$, as in the
  case of \emph{unconstrained} condensation, but with qualitative
  features more similar to the case of \emph{constrained} condensation
  with extensive mass: \eb{for $\delta > \delta_c$
(and for $\delta = \delta_c$ at large enough $\tilde{\rho}$)}
the marginal distribution $p(m)$ of the local
  mass has a secondary peak related to the condensate fraction, at
  variance with unconstrained case where $p(m)$ decays monotonously
  for increasing values of $m$.

The inclusion in the problem of the new parameter $\delta$,
  which characterizes the superextensive scaling $M \sim N^{1+\delta}$
  of the total mass, allowed us to draw the two-dimensional ($\alpha$,
  $\delta$) phase diagram shown in Fig.~\ref{fig1}. At variance with
  the two models usually studied in the literature, where condensation takes
  place \emph{either} for $\alpha<1$, without the constraint,
  \emph{or} for $\alpha>1$, with constrained extensive mass, in the
  case of a constrained superextensive mass condensation is found
  \emph{both} for $\alpha<1$ (\eb{when} $\delta>\delta_c$) \emph{and} for
  $\alpha>1$ (\eb{when} $\delta>0$).

More in detail, we have shown that
as soon as $\delta>0$, \eb{constrained} condensation occurs for any $\alpha >1$,
irrespective of the value of the generalized density $\tilde{\rho}$,
\emph{when the system is constrained to have a superextensive value of
  the mass}. This case is qualitatively similar to the case of an
extensive mass $M=\rho N$ with a large density $\rho$.  
For $\alpha <1$, \eb{a weak form of} condensation occurs if $\delta \ge
\delta_c(\alpha)$, \eb{in the sense that the participation ratio takes a nonzero value in the infinite $N$ limit.} Here, $1+\delta_c=1/\alpha$ is precisely the
scaling exponent of the mass in the unconstrained case. 
\eb{When $\delta > \delta_c$, condensation takes the form of a bump with vanishing relative width in the marginal distribution $p(m)$. It thus shares similarities with the standard condensation phenomenon. When $\delta = \delta_c$, only a pseudo-condensate with non-vanishing relative fluctuations appears, or the distribution $p(m)$ may even decay monotonously. This confirmed by the expression Eq.~(\ref{eq:Yk:superext:crit}) of $Y_k$ ($\alpha<1$ and constraint to superextensive mass with $\delta=\delta_c$), which differs from Eq.~(\ref{eq:Yk:constrained:extensive}) obtained in the case of $\alpha>1$ and a fixed extensive mass.
The situation is thus different from standard condensation, but the nonzero asymptotic value of the participation ratio indicates that some non-trivial phenomenon (that we call weak condensation) takes place.}


To conclude, we note that the qualitative idea that condensation
occurs when one imposes a total mass larger than the `natural' mass
the system would have in the unconstrained case remains valid: this is
always the case for $\alpha >1$ (both for extensive and superextensive
constraints), but it is also the case to some extent for $\alpha <1$,
where condensation is present for $\delta >\delta_c$. Yet, one has to
be aware that the notion of `natural mass' is not firmly grounded in
this case, and is just a heuristic concept associated to a typical
scaling $M \sim N^{1/\alpha}$ with the system size $N$.  One further
subtlety is whether condensation occurs or not on the transition
line. For the constrained case with an extensive mass, condensation
does not occur at the critical density $\rho=\rho_c$. In constrast,
\eb{a weak form of} condensation occurs at $\delta=\delta_c$
\eb{(see Sec.~\ref{super-extensive:critical})}, which may suggest a
discontinuous condensation transition as a function of $\delta$. But
for $\delta=\delta_c$ the condensation properties actually depend on
the generalized density $\tilde{\rho}$, see
Eq.~(\ref{eq:Yk:superext:crit}), so that \eb{this weak} condensation is actually
continuous (in the sense that $Y_k$ goes to zero when $\tilde{\rho}
\to 0$) if one looks on a finer scale in terms of $\tilde{\rho}$.\\

\acknowledgments{G.G. acknowledges Financial support from ERC Grant 
  No. ADG20110209.}

\medskip

\appendix

\section{Participation ratio for $M=\tilde{\rho} N^{1+\delta}$}

In this appendix, we provide some technical details on the evaluation of participation ratios for $\alpha >1$, in the cases $\delta > \delta_c$ and $\delta = \delta_c$.

\subsection{Case $\delta > \delta_c$}

Considering the case $\alpha <1$ and $\delta > \delta_c$, we wish here to justify the approximation made to go from Eq.~(\ref{eq:mk:superext:supercrit}) to Eq.~(\ref{eq:mk:superext:supercrit2}) in the evaluation of $\langle m^k \rangle$.
Considering the integral appearing in Eq.~(\ref{eq:mk:superext:supercrit}) as well as its approximation, one can write, setting $V_0 = M/N^{1/\alpha}$
\bea \label{eq:int:app3}
&& \int_0^{V_0} dv \,G(v) -\int_0^{V_0} dv \, G(v)
\left( 1-\frac{v}{V_0} \right)^{k-1-\alpha} \\ \nonumber
&& \qquad \qquad = V_0 \int_0^1 du\, G(V_0u)\, [1-(1-u)^{k-1-\alpha}] \\ \nonumber
&& \qquad \qquad \approx \frac{A}{V_0^{\alpha}} \int_0^1 \frac{du}{u^{1+\alpha}} \,[1-(1-u)^{k-1-\alpha}] \\ \nonumber
\eea
where we have used the change of variable $v=V_0 u$ as well as the asymptotic behavior of the function $G(v)$.
Assuming $k-\alpha>0$, the last integral in Eq.~(\ref{eq:int:app3}) converges, so that the difference of the two integrals in the lhs of Eq.~(\ref{eq:int:app3}) indeed converges to $0$ when $N \to \infty$, since $V_0 \to \infty$ in this limit.

\subsection{Case $\delta = \delta_c$}

We discuss here the asymptotic, small $\tilde{\rho}$ and large $\tilde{\rho}$, behavior of the participation ratio $Y_k$ in the case $\delta = \delta_c$. 

Let us start by the small $\tilde{\rho}$ regime.
As discussed in the main text, $Y_k$ can be obtained by an inverse Laplace transform, see Eq.~(\ref{eq:Yk:inverse:laplace}).
Introducing $\psi(\tilde{\rho})=\tilde{\rho}^{k-1-\alpha} Y_k(\tilde{\rho})$, the Laplace transform
$\hat{\psi}(s)$ is given by
\be
\hat{\psi}(s) = \frac{\Gamma(k-\alpha)}{s^{k-\alpha}} \, e^{-as^{\alpha}}
\ee
Taking the inverse Laplace transform, one has
\be \label{inv:laplace:eq1}
\psi(\tilde{\rho}) = \frac{1}{2\pi i}
\int_{s_0-i\infty}^{s_0+i\infty} ds \,\frac{\Gamma(k-\alpha)}{s^{k-\alpha}} \, e^{-as^{\alpha} + s\tilde{\rho}} .
\ee
In order to see whether this integral can be performed through a saddle-point evaluation, we note that balancing the two terms in the argument of the exponential leads to $as^{\alpha} \sim s\tilde{\rho}$, which results in $s \sim \tilde{\rho}^{-1/(1-\alpha)}$, eventually leading to $-as^{\alpha} + s\tilde{\rho} \sim \tilde{\rho}^{-\alpha/(1-\alpha)}$
(note that the algebraic prefactor in front of the exponential does not change the location of the saddle-point).
The argument can then be made sharper using the change of variable
$s=z/\tilde{\rho}^{1/(1-\alpha)}$, yielding
\be
-as^{\alpha} + s\tilde{\rho} = \tilde{\rho}^{-\alpha/(1-\alpha)} (-az^{\alpha} + z) .
\ee
In this form, a diverging prefactor is obtained when $\tilde{\rho} \to 0$, so that a saddle-point calculation can indeed be performed in this limit.
Defining $\phi(z)=-az^{\alpha} + z$, the saddle-point $z_0$ is obtained for
$\phi'(z_0)=0$, yielding $z_0=(\alpha a)^{1/(1-\alpha)}$.
Choosing $s_0=z_0/\tilde{\rho}^{1/(1-\alpha)}$ in the integral (\ref{inv:laplace:eq1}), and setting $z=z_0 + iy$, one can write in the small $\tilde{\rho}$ limit
\bea \nonumber
\psi(\tilde{\rho}) &\approx& \frac{\Gamma(k-\alpha)}{2\pi s^{k-\alpha}}
\int_{-\infty}^{\infty} dy \, \exp\left( \tilde{\rho}^{\frac{\alpha}{\alpha-1}} [\phi(z_0)-\phi''(z_0)\frac{y^2}{2}] \right)\\
&\approx& \frac{\Gamma(k-\alpha)}{z_0^{k-\alpha} \sqrt{2\pi \phi''(z_0)}}
\, \tilde{\rho}^{\lambda} \exp\left( \phi(z_0) \, \tilde{\rho}^{-\frac{\alpha}{1-\alpha}} \right)
\eea
where $\lambda$ is given in Eq.~(\ref{eq:def:lambda}), thus recovering
Eq.~(\ref{eq:Yk:crit:small:rho}).

We now turn to the computation in the large $\tilde{\rho}$ limit. We have seen that the inverse Laplace transform cannot be computed through a saddle-point evaluation in this limit. We thus come back to Eq.~(\ref{eq:Yk:superext:crit}) and rewrite it as
\be \label{eq:Yk:superext:crit:app}
Y_k = \int_0^{\tilde{\rho}} dv \, \left(1-\frac{v}{\tilde{\rho}}\right)^{k-1-\alpha} H(v)
\ee
Since the integral $\int_0^{\infty} dv \, H(v)$ converges, one can approximate for large $\tilde{\rho}$ the factor $(1-v/\tilde{\rho})^{k-1-\alpha}$ in Eq.~(\ref{eq:Yk:superext:crit:app}) by $1$, assuming $k-1-\alpha >0$.
Hence, again for $\tilde{\rho} \to \infty$,
\be
Y_k \approx \int_0^{\tilde{\rho}} dv \, H(v) \to \int_0^{\infty} dv \, H(v) = 1
\ee
The correction to $Y_k=1$ can be computed as follows:
\be \label{Yk:large:rho:corr}
1-Y_k = \int_{\tilde{\rho}}^{\infty} dv \, H(v)
+ \int_0^{\tilde{\rho}} dv \, H(v) \left[ 1- \left( 1-\frac{v}{\tilde{\rho}} \right)^{k-1-\alpha}  \right]
\ee
The first integral in Eq.~(\ref{Yk:large:rho:corr}) is easily evaluated for large $\tilde{\rho}$ as
\be \label{eq:Hv:large:rho}
\int_0^{\tilde{\rho}} dv \, H(v) \approx \frac{A}{\alpha \tilde{\rho}^{\alpha}} .
\ee
The second integral in Eq.~(\ref{Yk:large:rho:corr}) can be rewritten with the change of variable $v=\tilde{\rho} u$ as
\bea \label{eq:int:app1}
&&\tilde{\rho} \int_0^1 du \, H(\tilde{\rho} u) \,[1-(1-u)^{k-1-\alpha}] \\ \nonumber
&& \qquad \qquad \qquad \approx \frac{A}{\tilde{\rho}^{\alpha}} \int_0^1 \frac{du}{u^{1+\alpha}} [1-(1-u)^{k-1-\alpha}] \equiv  \frac{A \, I}{\tilde{\rho}^{\alpha}}
\eea
where we have denoted as $I$ the last integral, and where we have used the asymptotic behavior of $H(\rho)$ given in Eq.~(\ref{eq:asympt:Hrho}).
Assuming $k-1-\alpha >0$ (which is consistent since $k$ is in most cases of interest an integer $>1$), the integral $I$ can be computed through an integration by part, leading to
\be \label{eq:int:app2}
I = -\frac{1}{\alpha} + \frac{k-1-\alpha}{\alpha}
\, \frac{\Gamma(1-\alpha)\Gamma(k-1-\alpha)}{\Gamma(k-2\alpha)}
\ee
where we have also used the standard result
\be
\int_0^1 du\, u^{\mu-1} (1-u)^{\nu-1} = \frac{\Gamma(\mu) \Gamma(\nu)}{\Gamma(\mu+\nu)}
\ee
for $\mu, \, \nu >0$.
Then combining Eqs.~(\ref{Yk:large:rho:corr}), (\ref{eq:Hv:large:rho}), (\ref{eq:int:app1}) and (\ref{eq:int:app2}),
one eventually obtains Eq.~(\ref{eq:Yk:crit:large:rho}).

\end{document}